\documentclass[preprint,12pt]{elsarticle}
\usepackage{amssymb,amsfonts,amsmath,amsthm,enumerate,latexsym,amsthm,yhmath}
\usepackage{url}
\usepackage{graphicx}
\usepackage{tikz}
\usetikzlibrary{snakes,shapes}
\tikzstyle{tre}=[circle,draw,minimum size=3mm]
\tikzstyle{btre}=[circle,draw,minimum size=4.5mm]
\newcommand{\etq}[1]{%
\draw (#1) node {\tiny $#1$};}

\renewcommand{\leq}{\leqslant}
\renewcommand{\geq}{\geqslant}

\newcommand{\pathgr}{\!\rightsquigarrow\!{}}
\newcommand{\TT}{\mathcal{T}}
\newcommand{\FF}{\mathcal{F}}

 \theoremstyle{plain}
 \newtheorem{theorem}{Theorem}
 \newtheorem{lemma}[theorem]{Lemma}
 \newtheorem{corollary}[theorem]{Corollary}
 \newtheorem{proposition}[theorem]{Proposition}
 \theoremstyle{definition}

\newcommand{\ZZ}{\mathbb{Z}}

\begin{document}
\begin{frontmatter}
\title{The mean value of the squared path-difference distance for rooted phylogenetic trees}

\author[uib]{Arnau Mir}
\ead{arnau.mir@uib.es}
\author[uib]{Francesc Rossell\'o\corref{cor1}}
\ead{cesc.rossello@uib.es}
\cortext[cor1]{Corresponding author}
\address[uib]{Research Institute of Health Science (IUNICS) and Department of Mathematics and
  Computer Science,
  University of the Balearic Islands,
  E-07122 Palma de
  Mallorca, Spain}

\begin{abstract}
The path-difference metric is one of the oldest distances for the comparison of fully resolved phylogenetic trees, but its statistical properties are still quite unknown. In this paper we compute the mean value of the square of the path-difference metric between two fully resolved rooted phylogenetic trees with $n$ leaves, under the uniform distribution. This complements previous work by Steel and Penny, who computed this mean value for fully resolved unrooted phylogenetic trees.
\end{abstract}
\end{frontmatter}

\section{Introduction}

The definition and study of metrics for the comparison of rooted phylogenetic trees is a classical problem in phylogenetics \cite[Ch.~30]{fel:04}, motivated by the need to compare alternative phylogenetic trees for a given set of organisms obtained from different
datasets  or using  different reconstruction algorithms  \cite{hoef:05}. Other applications of these metrics include the assessment of phylogenetic tree reconstruction methods \cite{willcliff:taxon71} and the definition of search-by-similarity procedures  on databases~\cite{page:2005}.  

Many metrics for the
comparison of rooted phylogenetic trees on the same set of taxa have been proposed so far. Some of the first such metrics, defined around 40 years ago, were based
on the comparison of the vectors of lengths of (undirected) paths connecting pairs of
taxa in the corresponding trees. These metrics comprise, for instance, the euclidean distance between these
vectors  \cite{farris:sz69,farris:sz73},
the Manhattan distance between them \cite{willcliff:taxon71},
or the correlation between them \cite{phipps:sz71}.  Similar
metrics have also been defined for unrooted
phylogenetic trees \cite{bluis.ea:2003,puigbo.ea:07,distributions}. 
Let us point out here that, in the rooted case, these metrics satisfy the separation axiom of metrics (distance 0 means isomorphism) only for \emph{fully resolved}, or \emph{binary}, phylogenetic trees, and hence they are metrics, in the actual mathematical sense of the term, only in this case; cf.~ \cite{cardona.ea:08a}. In the unrooted case, they are metrics for arbitrary trees.

In contrast with other metrics \cite{Bryant.Steel:09,Hendy.ea:84,Steel:88,distributions}, and despite their tradition and popularity, the statistical properties of these path-lengths based metrics are mostly unknown. For instance, the diameter of none of these metrics (either in the rooted or in the unrooted case) is known yet. Steel and Penny \cite{distributions} studied, among others, the distribution of one of these distances for unrooted trees: the one defined through the euclidean distance between path-lengths vectors, which these authors called the \emph{path-difference metric} (other published names for this metric are the \emph{cladistic difference} \cite{farris:sz69} and, generically, a \emph{nodal distance} \cite{bluis.ea:2003,puigbo.ea:07}).   In the aforementioned paper,  Steel and Penny computed the mean  value of the square  of this path-difference metric for fully resolved unrooted trees. The knowledge of this mean value is useful in the assessment of a comparison of two trees through this metric, because it ``provides an indication as to whether or not this measured similarity could have come about by chance" \cite{distributions}.

In this paper we compute the mean  value of the square  of the path-difference metric for fully resolved rooted phylogenetic trees with $n$ leaves. Although the raw argument underlying our computation is the same as in Steel and Penny's paper, the details in the rooted case are much harder than in the unrooted case, because of the asymmetric role of the root. We have proved that this mean value grows in $O(n^3)$; more specifically, it is
$$
2\binom{n}{2}\left(4(n-1)+2-\frac{2^{2(n-1)}}{\binom{2(n-1)}{n-1}}-\left(\frac{2^{2(n-1)}}{\binom{2(n-1)}{n-1}}\right)^2\right).
$$
This turns out to be the mean value obtained by Steel and Penny for unrooted phylogenetic trees, but with $n+1$ leaves. A similar relationship between combinatorial values for rooted and unrooted phylogenetic trees arises in other problems; for instance, a simple argument shows that  the number of rooted phylogenetic trees with $n$ leaves is the number of unrooted phylogenetic trees with $n+1$ leaves   \cite[Ch.~3]{fel:04}; also, as we shall see in this paper (Corollary \ref{cor:meand}), the mean value of the length of the undirected path between two given leaves in a rooted phylogenetic tree with $n$ leaves is equal to the corresponding mean value for unrooted phylogenetic trees. But we have not been able to find a clever argument that proves directly this relationship between the mean values of the squared path-difference metric, or
of the path-length between two leaves,  in the rooted and unrooted cases, and thus we have needed to compute them.

\section{Preliminaries}
\label{sec:prel}

\subsection{Phylogenetic trees}
In this paper, by a  \emph{phylogenetic tree} on a set $S$ of taxa we mean a \emph{fully resolved}, or
\emph{binary} (that is, with   all its internal nodes of out-degree 2), rooted tree  with its leaves bijectively labeled in  the set $S$.   To simplify the language, we shall always identify a leaf of a phylogenetic tree with its label.  We shall also use the term \emph{phylogenetic tree with $n$ leaves} to refer to a phylogenetic tree on a given set of $n$ taxa, when this set  is known or nonrelevant. 

We shall represent a path
from $u$ to $v$ in a phylogenetic tree $T$
by $u \pathgr v$.  Whenever there exists a path
$u\pathgr v$, we shall say that $v$ is a 
\emph{descendant} of $u$ and also that $u$ is an 
\emph{ancestor} of $v$. 
Given a node $v$ of a phylogenetic tree $T$, the \emph{subtree of $T$ rooted at $v$} is the subgraph of $T$ induced on the set of descendants of $v$. It is a phylogenetic tree on the set of descendant leaves of $v$, and with root this node $v$.

The \emph{lowest common ancestor} (LCA) of a pair of nodes
$u,v$ of a phylogenetic tree $T$, in symbols
$LCA_T(u,v)$, is the unique common ancestor of them that is
a descendant of every other common ancestor of them.  The \emph{path difference} $d_T(u,v)$ between two nodes $u$ and $v$ is the sum of the lengths of the paths $LCA_T(u,v)\pathgr u$ and 
 $LCA_T(u,v)\pathgr v$; equivalently, it is the length of the only path connecting $u$ and $v$ in the undirected tree associated to $T$.
 It is well-known (for a proof, see \cite{cardona.ea:08a}) that the vector of path differences $d(T)=\big(d_T(i,j)\big)_{1\leq i<j\leq n}$ between all pairs of leaves characterizes up to isomorphism a phylogenetic tree with $n$ leaves: this property is false if we remove the binarity assumption on the trees.
 
Let $\TT_n$ be the set of (isomorphism classes of) phylogenetic trees with $n$ leaves. It is well known \cite[Ch. 3]{fel:04} that $|\TT_1|=1$ and, for every $n\geq 2$,
$$
|\TT_n|=(2n-3)!!=(2n-3)(2n-5)\cdots 3 \cdot 1.
$$
An \emph{ordered $m$-forest} on a set $S$  is an ordered sequence of $m$ phylogenetic trees $(T_1,T_2,\ldots,T_m)$, each $T_i$ on a set $S_i$ of taxa, such that these sets $S_i$ are pairwise disjoint and their union is $S$. 
Let $\FF_{m,n}$ be the set of (isomorphism classes of) ordered $m$-forests on any given set $S$ with $|S|=n$. The cardinal of $\FF_{m,n}$ is computed (although not explicitly) along the proof of Theorem 3 in \cite{distributions}.

\begin{lemma}\label{lem:|F|}
For every $m\geq 1$, $|\FF_{m,m}|=m!$ and
$$
|\FF_{m,n}|= \frac{m (n!) \prod_{l=1}^{n-m-1} (n+l)}{(2 (n-m))!!}=\frac{(2n-m-1)!m}{(n-m)!2^{n-m}}
% \frac{m (n!) \prod_{l=1}^{n-m-1} (n+l)}{(2 (n-m))!!}
 \quad\mbox{for every $n>m$}.
$$
\end{lemma}

\begin{proof}
The
exponential generating function for the number of rooted phylogenetic trees with $n$ leaves is
$B(x)=1-\sqrt{1-2x}$.
Then, the exponential  generating function for the number of ordered forests consisting of a given number of trees (marked by the variable $y$) and a given global number of leaves (marked by the variable $y$) is
$$
F(x,y)=\sum_{m\geq 1} y^mB(x)^m= \frac{1}{1-yB(x)}-1.
$$
This implies that the number $|\FF_{m,n}|$ of ordered $m$-forests on a set of $n$ leaves is equal to 
$\frac{\partial^{n}}{\partial x^{n}} \left( B(x)^{m}\right) \big|_{{x=0}}$.
This derivative can be easily computed, yielding the values given in the statement.
\end{proof}
 
\subsection{Hypergeometric functions}
The (\emph{generalized}) \emph{hypergeometric function} ${}_pF_q$ is defined  \cite{Bayley} as 
$$
{}_pF_q\bigg(\begin{array}{rrr} a_1,&\ldots, &a_p \\[-0.5ex] b_1,& \ldots,  &b_q\end{array};z\bigg)=\sum_{k\geq 0} \frac{(a_1)_k\cdots (a_p)_k}{(b_1)_k\cdots (b_q)_k}\cdot \frac{z^k}{k!},
$$ 
where $(a)_k := a\cdot (a+1)\cdots (a+k-1)$ .

The following lemmas  will be used in the next section.

\begin{lemma}\label{lem:prel3}
$$
{}_2F_1 \bigg(\begin{array}{rr} n-1,&2-n \\[-0.5ex] -n& \end{array};\dfrac{1}{2}\bigg)= \frac{2^{n-1}}{n}.
$$
\end{lemma}

\begin{proof}
To compute the value of ${}_2F_1 \bigg(\begin{array}{rr} n-1,&2-n \\[-0.5ex] -n& \end{array};\dfrac{1}{2}\bigg)$ we shall use Formula 15.1.26 in \cite{HMM} (see also  \url{http://functions.wolfram.com/07.23.03.0028.01}):
$$
 {}_2F_1 \bigg(\begin{array}{rr} a,&1-a \\[-0.5ex] c& \end{array};\frac{1}{2}\bigg)=\frac{2^{1-c} \sqrt{\pi} \Gamma(c)}{\Gamma\left(\frac{a+c}{2}\right) \Gamma\left(\frac{c-a+1}{2}\right)}.
$$
We cannot apply this expression to $a=n-1$ and $c=-n$, because $\Gamma (-n)=\infty$. So, instead, we use a standard pass to limit argument:
$$
\begin{array}{rl}
\displaystyle{}_2F_1 \bigg(\begin{array}{rr} n-1,&2-n \\[-0.5ex] -n& \end{array};\frac{1}{2}\bigg) & \displaystyle =  \lim_{\varepsilon\to 0} {}_2F_1 \bigg(\begin{array}{rr} n-1,&2-n \\[-0.5ex] -n+\varepsilon& \end{array};\frac{1}{2}\bigg) \\ & \displaystyle =\lim_{\varepsilon\to 0} \frac{2^{1+n-\varepsilon} \sqrt{\pi} \Gamma(-n+\varepsilon)}{\Gamma\left(\frac{\varepsilon-1}{2}\right) \Gamma\left(\frac{2-2n+\varepsilon}{2}\right)} =\frac{2^{n-1}}{n}.
\end{array}
$$
\end{proof}

\begin{lemma}\label{lem:prel2}
$$
 {}_3F_2 \bigg(\begin{array}{rrr} 1-n,&2-n,&n-1 \\[-0.5ex] -n,&-n& \end{array};\frac{1}{2}\bigg)  = \frac{2^{n-1}}{n^2} \left(-1 +\frac{ (2n-1)!!}{2^{n-2} (n-1)!} \right),
$$
\end{lemma}

\begin{proof}
The hypergeometric series $ {}_3F_2 \bigg(\begin{array}{rrr} 1-n,&2-n,&n-1 \\[-0.5ex] -n,&-n& \end{array};\dfrac{1}{2}\bigg) $ can be written as a function of the hypergeometric function $ {}_2F_1 $ as follows:\footnote{See \url{http://functions.wolfram.com/07.27.03.0118.01}}
\begin{equation}
\begin{array}{rl}
{}_3F_2 \bigg(\begin{array}{rrr} 1-n,&2-n,&n-1 \\[-0.5ex] -n,&-n& \end{array};\dfrac{1}{2}\bigg) = & {}_2F_1 \bigg(\begin{array}{rr} n-1,&2-n \\[-0.5ex] -n& \end{array};\dfrac{1}{2}\bigg)\\ & \displaystyle\ -\frac{(n-1)(n-2)}{2 n^2} {}_2F_1 \bigg(\begin{array}{rr} n,&3-n \\[-0.5ex] 1-n& \end{array};\dfrac{1}{2}\bigg).
\end{array}
\label{SIMP3F2}
\end{equation}
We already know from the previous lemma that
${}_2F_1 \bigg(\begin{array}{rr} n-1,&2-n \\[-0.5ex] -n& \end{array};\dfrac{1}{2}\bigg)=\dfrac{2^{n-1}}{n}$. It remains to compute ${}_2F_1 \bigg(\begin{array}{rr} n,&3-n \\[-0.5ex] 1-n& \end{array};\dfrac{1}{2}\bigg)$. To do it, we shall use the following formula:\footnote{See
\url{http://functions.wolfram.com/07.23.03.0030.01}.}
$$
 {}_2F_1 \bigg(\begin{array}{rr} a,&3-a \\[-0.5ex] c& \end{array};\frac{1}{2}\bigg)=\frac{2^{3-c} \sqrt{\pi} \Gamma(c)}{(a-1)(a-2)}\left( \frac{c-2}{\Gamma\left(\frac{a+c}{2}-1\right) \Gamma\left(\frac{c-a+1}{2}\right)} - \frac{2}{\Gamma\left(\frac{a+c-3}{2}\right)\Gamma\left(\frac{c-a}{2}\right)}\right).
$$
Again, we cannot apply this formula  to $a=n-1$ and $c=-n$, and thus we use a pass to limit argument: 
$$
\begin{array}{l}
\displaystyle {}_2F_1 \bigg(\begin{array}{rr} n,&3-n \\[-0.5ex] 1-n& \end{array};\frac{1}{2}\bigg)  =  \lim_{\varepsilon\to 0}  {}_2F_1 \bigg(\begin{array}{rr} n,&3-n \\[-0.5ex] 1-n+\varepsilon& \end{array};\frac{1}{2}\bigg) \\ \displaystyle \qquad =  \lim_{\varepsilon\to 0}  \frac{2^{2+n-\varepsilon} \sqrt{\pi}\Gamma(1-n+\varepsilon) }{(n-1)(n-2)}\left(\frac{(-n-1-\varepsilon)}{\Gamma\left(\frac{\varepsilon-1}{2}\right)\Gamma\left(1-n+\frac{\varepsilon}{2}\right)} -\frac{2}{\Gamma\left(\frac{\varepsilon -2}{2}\right)\Gamma\left(\frac{1+\varepsilon-2n}{2}\right)}\right)
\\  \displaystyle \qquad = \frac{2^{2+n} \sqrt{\pi} }{(n-1)(n-2)} \left( \lim_{\varepsilon\to 0} \frac{(-n-1-\varepsilon)\Gamma(1-n+\varepsilon)}{\Gamma\left(\frac{\varepsilon-1}{2}\right)\Gamma\left(1-n+\frac{\varepsilon}{2}\right)} - \lim_{\varepsilon\to 0} \frac{2\Gamma(1-n+\varepsilon)}{\Gamma\left(\frac{\varepsilon -2}{2}\right)\Gamma\left(\frac{1+\varepsilon-2n}{2}\right)}\right) \\ \displaystyle \qquad = \frac{2^{2+n} \sqrt{\pi} }{(n-1)(n-2)} \left(\frac{(n+1)}{4\sqrt{\pi}} - \frac{(-1)^{n+2} \binom{-1/2}{n} n!}{(n-1)! \sqrt{\pi}}\right)\\ \displaystyle \qquad= \frac{2^{2+n} \sqrt{\pi} }{(n-1)(n-2)} \left(\frac{(n+1)}{4} -\frac{(2n-1)!!}{(n-1)! 2^n}\right).
\end{array}
$$

Replacing ${}_2F_1 \bigg(\begin{array}{rr} n-1,&2-n \\[-0.5ex] -n& \end{array};\dfrac{1}{2}\bigg)$ and ${}_2F_1 \bigg(\begin{array}{rr} n,&3-n \\[-0.5ex] 1-n& \end{array};\dfrac{1}{2}\bigg) $  in equation~(\ref{SIMP3F2}) by their values given above, we obtain
$$
\begin{array}{rl}
\displaystyle{}_3F_2 \bigg(\begin{array}{rrr} 1-n,&2-n,&n-1 \\[-0.5ex] -n,&-n& \end{array};\frac{1}{2}\bigg) & \displaystyle = \frac{2^{n-1}}{n}-\frac{2^{n+2}}{2 n^2}\left(\frac{(n+1)}{4}-\frac{(2n-1)!!}{(n-1)! 2^n}\right)\\ & \displaystyle = \frac{2^{n-1}}{n^2}\left(-1+\frac{(2n-1)!!}{2^{n-2} (n-1)!}\right).
\end{array}
$$
as we claimed.
\end{proof}

\begin{lemma}\label{lem:prel1}
For every real numbers $a,b$,
$$
\begin{array}{l}
{}_4F_3 \bigg(\begin{array}{rrrr} 1,&a,&a+1/2,&b \\[-0.5ex] 2,& 2a, & b+1/2& \end{array};1\bigg)\\ \qquad\qquad\qquad\quad \displaystyle =\frac{(2b-1)}{(a-1)(b-1)} \left(-1 + {}_3F_2  \bigg(\begin{array}{rrr} a-1,&a-1/2,&b-1 \\[-0.5ex] 2a-1,& b-1/2& \end{array};1\bigg)\right).
\end{array}
$$
\end{lemma}

\begin{proof}
By definition,
$$
\begin{array}{rl}
\displaystyle {}_4F_3  \bigg(\begin{array}{rrrr} 1,&a,&a+1/2,&b \\[-0.5ex] 2,& 2a, & b+1/2& \end{array};1\bigg)& \displaystyle=\sum_{k\geq 0} \frac{k!(a)_k (a+1/2)_k (b)_k}{(k+1)! (2a)_k (b+1/2)_k}\cdot \frac{1}{k!}\\
& \displaystyle= \sum_{k\geq 1} \frac{(a)_{k-1} (a+1/2)_{k-1} (b)_{k-1}}{k! (2a)_{k-1} (b+1/2)_{k-1}}=(*).
\end{array}
$$
Taking into account that 
\begin{eqnarray*}
(a)_{k-1} &= &\frac{(a-1)_k}{a-1},\ (a+1/2)_{k-1}=\frac{(a-1/2)_k}{a-1/2},\ (b)_{k-1}=\frac{(b-1)_k}{b-1},\\ (2a)_{k-1} &= &\frac{(2a-1)_k}{2a-1},\ (b+1/2)_{k-1}=\frac{(b-1/2)_k}{b-1/2},
\end{eqnarray*}
the expression (*) can be written as
$$
\begin{array}{rl}
(*) & = \displaystyle
 \sum_{k\geq 1} \frac{(a-1)_{k} (a-1/2)_{k} (b-1)_{k} (2a-1)(b-1/2)}{(a-1)(a-1/2)(b-1) (2a-1)_{k} (b-1/2)_{k}}\cdot \frac{1}{k!}\\
  & = \displaystyle
\frac{(2b-1)}{(a-1)(b-1)} \left(-1 + {}_3F_2  \bigg(\begin{array}{rrr} a-1,&a-1/2,&b-1 \\[-0.5ex] 2a-1,& b-1/2& \end{array};1\bigg)\right)
\end{array}
$$
yielding the formula in the statement.
\end{proof}

\section{Mean total areas}
 
For every $s\in \ZZ^+$, the \emph{total $s$-area} of a phylogenetic tree $T$  is 
$$
D^{(s)}(T)=\displaystyle\sum_{1\leq i<j\leq n} d_T(i,j)^s.
$$
This value (or, rather, its $s$-th root) measures the total amount of evolutive history captured by the phylogenetic tree.
 Let 
$$
\mu(D^{(s)})_n=\frac{\sum_{T\in \TT_n} D^{(s)}(T)}{|\TT_n|}
$$
be the mean value  of $D^{(s)}(T)$ for $T\in \TT_n$ under the uniform distribution on $\TT_n$. In this section we compute $\mu(D^{(1)})_n$ and $\mu(D^{(2)})_n$.
To simplify the notations, for every $s\in \ZZ^+$ let
$$
S_n^{(s)}=\sum_{T\in \TT_n} d_T(1,2)^s.
$$

\begin{lemma}\label{lem:eq1}
For every $s\in \ZZ^+$ and for every $1\leq i<j\leq n$,
$$
\sum_{T\in \TT_n} d_T(i,j)^s=S_n^{(s)}.
$$
\end{lemma}

\begin{proof}
Let $\sigma_{i,j}$ be the involutive permutation that interchanges 1 and $i$, and $2$ and $j$ and leaves the other elements fixed and, for every $T\in \TT_n$, let $T_{\sigma_{i,j}}$ be the phylogenetic tree obtained by applying to the leaves in $T$ the permutation $\sigma_{i,j}$. On the one hand, it is clear that
$d_T(i,j)=d_{T_{\sigma_{i,j}}}(1,2)$, and, on the other hand, since the mapping
$\TT_n  \to  \TT_n$ defined by
$T  \mapsto  T_{\sigma_{i,j}}$
is bijective, we have the equality of multisets
$$
\big\{d_T(1,2)\mid T\in \TT_n\big\}=\big\{d_{T_{\sigma_{i,j}}}(1,2)\mid T\in \TT_n\big\}.
$$
Combining these two observations we obtain
$$
\sum_{T\in \TT_n} d_T(i,j)^s=\sum_{T\in \TT_n} d_{T_{\sigma_{i,j}}}(1,2)^s=\sum_{T\in \TT_n} d_T(1,2)^s.
$$
\end{proof}
 
\begin{corollary}\label{lem:mu->Sn}
For every $n\geq 2$ and for every $s\in \ZZ^+$,
$$
\mu(D^{(s)})_n=\binom{n}{2}\frac{S_n^{(s)}}{(2n-3)!!}.
$$
\end{corollary}

\begin{proof}
Using the previous lemma,
$$
\mu(D^{(s)})_n=\frac{\sum_{T\in \TT_n} D^{(s)}(T)}{|\TT_n|}
=\frac{\sum_{1\leq i<j\leq n}\sum_{T\in \TT_n} d_T(i,j)^s}{(2n-3)!!}=
\frac{\binom{n}{2}\sum_{T\in \TT_n} d_T(1,2)^s}{(2n-3)!!}.
$$
\end{proof}

For every $i=1,\ldots,n-1$, let $c_i$ be the cardinal of the set 
$$
\{ T\in \TT_n\mid d_T(1,2)=i\}.
$$
 Then, $S_n^{(s)} =\sum_{i=2}^{n} i^s c_i$.
Our first goal is to find a suitable expression for these coefficients $c_i$.

\begin{proposition}\label{prop:ci}
$\displaystyle c_i = \frac{(i-1) (2n-i-2)!}{(2(n-i))!!}$.
\end{proposition}

\begin{proof}
Let $T\in \TT_n$ be any tree such that $d_T(1,2)=i$; to simplify the notations, let us denote by $x$ the node $LCA_T(1,2)$. Then, on the one hand, 
the paths $x\pathgr 1$ and  $x\pathgr 2$ have, respectively, $j$ and $i-2-j$ intermediate nodes, for some $j=0,\ldots,i-2$, and each such intermediate node is the parent of the root of a rooted subtree of $T$. Let $\{i_1,\ldots,i_k\}$ be the union of the (pairwise disjoint) sets of leaves of these subtrees: notice that $i-2\leq k\leq n-2$, because each subtree has some leaf and the leaves $1,2$ cannot belong to these subtrees.
On the other hand,  $x$ is the leaf of the phylogenetic tree $T_0$ with leaves
$\big(\{1,\ldots,n\}\setminus\{1,2,i_1,\ldots,i_k\}\big)\cup \{x\}$ obtained by collapsing the subtree of $T$ rooted at $x$ into a single leaf $x$.

So, the tree $T$ is determined by a subset $\{i_1,\ldots,i_k\}$ of $\{1,\ldots, n\}$, with $i-2\leq k\leq n-2$, a phylogenetic tree $T_0$  on $\big(\{1,\ldots,n\}\setminus\{1,2,i_1,\ldots,i_k\}\big)\cup \{x\}$ (and hence with $n-k-1$ leaves), an ordered $(i-2)$-forest  $(T_1,\ldots,T_{i-2})$ on  $\{i_1,\ldots,i_k\}$, and an index $j\in\{0,1,\ldots,i-2\}$. The tree $T$ is obtained by starting in the leaf $x$ of $T_0$ two new paths $(x,v_j,\ldots,v_1,1)$ and $(x,v_{j+1},\ldots,v_{i-2},2)$ of lengths $j+1$ and $i-j-1$, respectively, and then adding to each intermediate node $v_l$ in these paths an arc with head the root of the tree $T_l$ (cf.~Fig.~\ref{fig:forests}).

\begin{figure}[htb]
\begin{center}
\begin{tikzpicture}[thick,>=stealth,scale=0.4]
\draw(0,0) node[btre] (x) {}; \etq x
\draw (x)--(10,0)--(5,3)--(x);
\draw(7,-0.6) node  {\scriptsize $\{1,\ldots,n\}\setminus\{1,2,i_1,\ldots,i_k\}$};
\draw(5,1) node  {\scriptsize $T_0$};
\draw(-3,-2) node[btre] (v_{j}) {}; \etq {v_{j}}
\draw(-7.5,-5) node[btre] (v_{1}) {}; \etq {v_{1}}

\draw(v_{j})--(-4.125,-2.75);
\draw(-4.8,-3.2) node  {.};
\draw(-5.25,-3.5) node  {.};
\draw(-5.7,-3.8) node  {.};
\draw(v_{1})--(-6.375,-4.25);

\draw(-10.5,-7) node[btre] (1) {}; \etq 1
\draw(3,-2) node[btre] (v_{j+1}) {}; \draw (v_{j+1}) node {\tiny $v_{j\!+\!1}$};
\draw(7.5,-5) node[btre] (v_{i-2}) {}; \draw (v_{i-2}) node {\tiny $v_{i\!-\!2}$};
\draw(v_{j+1})--(4.125,-2.75);
\draw(4.8,-3.2) node  {.};
\draw(5.25,-3.5) node  {.};
\draw(5.7,-3.8) node  {.};
\draw(v_{i-2})--(6.375,-4.25);
\draw(10.5,-7) node[btre] (2) {}; \etq 2
\draw(-2,-4) node[btre] (wj) {};
\draw (wj)--(-0.5,-7)--(-3.5,-7)--(wj);
\draw(-2,-6) node  {\scriptsize $T_j$};
\draw(-6.5,-7) node[btre] (w1) {};
\draw (w1)--(-5,-10)--(-8,-10)--(w1);
\draw(-6.5,-9) node  {\scriptsize $T_1$};
\draw(2,-4) node[btre] (wj+1) {};
\draw (wj+1)--(0.5,-7)--(3.5,-7)--(wj+1);
\draw(2,-6) node  {\scriptsize $T_{j+1}$};
\draw(6.5,-7) node[btre] (w2) {};
\draw (w2)--(5,-10)--(8,-10)--(w2);
\draw(6.5,-9) node  {\footnotesize $T_{i-2}$};
\draw (x)--(v_{j});
\draw (v_{1})--(1);
\draw (x)--(v_{j+1});
\draw (v_{i-2})--(2);
\draw (v_{j})--(wj);
\draw (v_{1})--(w1);
\draw (v_{i-2})--(w2);
\draw (v_{j+1})--(wj+1);

\end{tikzpicture}

\end{center}
\caption{\label{fig:forests} 
The structure of a tree $T$ with $d_T(1,2)=i$.}
\end{figure}
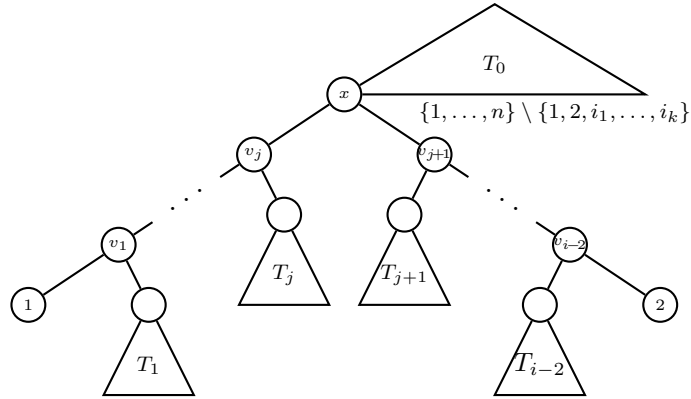

This shows that $c_i$ can be computed as
$$
\begin{array}{rll}
c_i & = \displaystyle\sum _{k=i-2}^{n-2}  \mbox{(number of ways  of choosing the $k$ nodes $i_1,\ldots,i_k$)}\\ & \qquad\qquad \cdot \mbox{(number of ordered $(i-2)$-forests  trees on $\{i_1,\ldots,i_k\}$)} \\
& \qquad\qquad \cdot \mbox{(number of ways of choosing $j$ between 0 and $i-2$)}\\
&  \qquad\qquad\cdot \mbox{(number of phylogenetic trees with $n-k-1$ leaves)}\\[2ex]
& = \displaystyle \sum_{k=i-2}^{n-2}  \binom{n-2}{k}\cdot |\FF_{i-2,k}|\cdot (i-1)\cdot (2n-2k-5)!!\\ & =
 \displaystyle (i-1)! \binom{n-2}{i-2}  (2n-2i-1)!! \\&  \qquad \displaystyle + (i-1) (i-2) \sum_{k=0}^{n-i-1} \binom{n-2}{k+i-1} \frac{(k+i-1)! \prod_{l=1}^k (k+i-1+l)}{(2(k+1))!!}  (2n-2k-2i-3)!!.
\end{array}
$$

Applying the \emph{hypergeometric series lookup algorithm} given in \cite[p. 36]{AeqB}, we obtain
$$
\begin{array}{l}
\displaystyle \sum_{k=0}^{n-i-1} \binom{n-2}{k+i-1} \frac{(k+i-1)! \prod_{l=1}^k (k+i-1+l)}{(2(k+1))!!} (2n-2k-2i-3)!!.\\ \displaystyle \qquad\qquad= \frac{1}{2} 	\binom{n-2}{i-1} (i-1)! (2n-2i-3)!!  \cdot {}_4F_3 \bigg(\begin{array}{rrrr} 1,&i/2,&i/2+1/2,&i-n+1 \\[-0.5ex] 2,& i, & i-n+3/2,& \end{array};1\bigg),
\end{array}
$$
and hence
$$
\begin{array}{rl}
c_i = & \displaystyle (i-1)! \binom{n-2}{i-2}  (2n-2i-3)!! \\ & \qquad\qquad \cdot \left( 2n-2i-1+\frac{1}{2} (i-2)(n-i)  {}_4F_3 \bigg(\begin{array}{rrrr} 1,&i/2,&i/2+1/2,&i-n+1 \\[-0.5ex] 2,& i, & i-n+3/2,& \end{array};1\bigg)\right).
\end{array}
$$
If we apply Lemma \ref{lem:prel1} with  $a=i/2$ and $b=i-n+1$, we obtain
\begin{eqnarray}\label{eq:ci-1}
c_i = (i-1)! \binom{n-2}{i-2} (2n-2i-1)!!  {}_3F_2 \bigg(\begin{array}{rrr} i/2-1,&i/2-1/2,&i-n \\[-0.5ex] i-1,& i-n+1/2& \end{array};1\bigg).
\end{eqnarray}
The value of $ {}_3F_2 \bigg(\begin{array}{rrr} i/2-1,&i/2-1/2,&i-n \\[-0.5ex] i-1,& i-n+1/2& \end{array};1\bigg)$ is computed in \cite[Form.~(2), p. 9]{Bayley}:
$$
 {}_3F_2 \bigg(\begin{array}{rrr} i/2-1,&i/2-1/2,&i-n \\[-0.5ex] i-1,& i-n+1/2& \end{array};1\bigg) = \frac{(i-2)! \Gamma(1/2+i-n) \Gamma (n-i/2) \Gamma (3/2-i/2)}{\Gamma (i/2) (n-2)! \Gamma (3/2+i/2-n) \Gamma(1/2)}.
$$
Now,  using that
$$
\Gamma (x)=(x-1)\Gamma(x-1),
$$
 we can write
\begin{eqnarray*}
\Gamma (3/2-i/2) & = & \frac{(-1)^{n-i} \prod_{k=1}^{n-i} (2k+i-3)}{2^{n-i}} \Gamma(3/2+i/2-n), \\
\Gamma (1/2) & = & \frac{(-1)^{n-i} \prod_{k=1}^{n-i} (2k-1)}{2^{n-i}} \Gamma(1/2+i-n), \\
\Gamma (n-i/2) & = & \frac{ \prod_{k=1}^{n-i} (2n-i-2k)}{2^{n-i}} \Gamma(i/2). 
\end{eqnarray*}
Then, using these formulas, the expression for $ {}_3F_2 \bigg(\begin{array}{rrr} i/2-1,&i/2-1/2,&i-n \\[-0.5ex] i-1,& i-n+1/2& \end{array};1\bigg)$ can be simplified, yielding
$$
 {}_3F_2 \bigg(\begin{array}{rrr} i/2-1,&i/2-1/2,&i-n \\[-0.5ex] i-1,& i-n+1/2& \end{array};1\bigg) = \frac{(i-2)! (2n-i-2)!}{(n-2)! (2n-2i-1)!! 2^{n-i}}.
$$
Replacing this expression in equation (\ref{eq:ci-1}), we finally obtain 
$$
c_i = \frac{(i-1) (2n-i-2)!}{(2(n-i))!!},
$$
as we claimed.
\end{proof}

\begin{proposition}\label{prop:muD1}
$S^{(1)}_n =2^{n-1} (n-1)!=(2 n-2)!!.$
\end{proposition}

\begin{proof}
By Proposition \ref{prop:ci} we know that
$$
S^{(1)}_n   =  \sum_{i=2}^{n} \frac{i (i-1) (2n-i-2)!}{ (2(n-i))!!} = \sum_{i=0}^{n-2} \frac{(i+2) (i+1)  (2n-i-4)!}{ (2 (n-i-2))!!} .
$$
If we compute this sum using again the algorithm given in \cite[p. 36]{AeqB}, we obtain
$$
S^{(1)}_n = \frac{2^{3-n} (2n-4)!}{(n-2)!} {}_2F_1 \bigg(\begin{array}{rr} 3,&2-n \\[-0.5ex] 4-2n& \end{array};2\bigg).
$$
Replacing in this equality ${}_2F_1 \bigg(\begin{array}{rr} 3,&2-n \\[-0.5ex] 4-2n& \end{array};2\bigg)$ by its definition, we obtain
$$
S^{(1)}_n = 2^{2-n} \sum_{k=0}^{n-2} \frac{(k+1) (k+2) (2n-k-4)!}{(n-k-2)!} 2^k = \sum_{k=0}^{n-2} \frac{(n-k-1) (n-k) (n+k-2)!}{k!} 2^{-k}.
$$
This sum can be computed using again the algorithm  given in \cite[p. 36]{AeqB}, yielding
$$
S^{(1)}_n = n! {}_2F_1 \bigg(\begin{array}{rr} n-1,&2-n \\[-0.5ex] -n& \end{array};1/2\bigg).
$$
By Lemma \ref{lem:prel3}, we conclude that
$$
S^{(1)}_n = \frac{n! 2^{n-1}}{n} =2^{n-1} (n-1)!=(2 n-2)!!,
$$
as we claimed.
\end{proof}

\begin{proposition}\label{prop:muD2}
$S_n^{(2)}  = 2 \cdot (2n-1)!!-(2n-2))!!.$
\end{proposition}

\begin{proof}
By Proposition \ref{prop:ci}, we have
$$
S_n^{(2)}=\sum_{i=2}^{n} i^2 c_i=\sum_{i=2}^{n} \frac{i^2 (i-1) (2n-i-2)!}{ (2(n-i))!!} = \sum_{i=0}^{n-2} \frac{(i+2)^2 (i+1)  (2n-i-4)!}{ (2 (n-i-2))!!} .
$$
Using again the algorithm  given in \cite[p. 36]{AeqB}, the value of the sum $S_n^{(2)}$ is:
\begin{eqnarray*}
S_n^{(2)} & = & \frac{2^{4-n} (2n-4)!}{(n-2)!} {}_3F_2 \bigg(\begin{array}{rrr} 3,&3,&2-n \\[-0.5ex] 2,&4-2n& \end{array};2\bigg)
=\sum_{k=0}^{n-2} \frac{(k+2)^2 (k+1) (2n-k-4)!}{(n-k-2)!} 2^{k-n+2} \\
& = & \sum_{k=0}^{n-2} \frac{(n-k)^2 (n-k-1) (n+k-2)!}{k!} 2^{-k}.
\end{eqnarray*}
Using once again the algorithm  given in \cite[p. 36]{AeqB}, the last sum can be computed as:
$$
S_n^{(2)}  = n\cdot n!  {}_3F_2 \bigg(\begin{array}{rrr} 1-n,&2-n,&n-1 \\[-0.5ex] -n,&-n& \end{array};\frac{1}{2}\bigg)
$$
By Lemma \ref{lem:prel2}, we obtain
$$
S_n^{(2)}  = (n-1)! 2^{n-1} \left(-1 +\frac{ (2n-1)!!}{2^{n-2} (n-1)!} \right)=2 \cdot (2n-1)!!-(2n-2))!!,
$$
as we claimed. \end{proof}

Applying Corollary \ref{lem:mu->Sn}, we obtain the following total areas.

\begin{corollary}
$$\displaystyle \mu(D^{(1)})_n=\binom{n}{2} \cdot\frac{(2n-2)!!}{(2n-3)!!},\qquad \mu(D^{(2)})_n=\displaystyle\binom{n}{2}\frac{2 (2n-1)!!-(2n-2)!!}{(2n-3)!!}.$$
\end{corollary}

$S_n^{(1)}$ can also be used to compute the mean value of the length of the undirected path between two given leaves in a phylogenetic tree.

\begin{corollary}\label{cor:meand}
For every $i,j\in \{1,\ldots,n\}$, $i\neq j$,
the mean value of $d_T(i,j)$ for $T\in \TT_n$ under the uniform distribution is
$$
\mu(d_T(i,j))_n=\frac{2^{2(n-1)}}{\binom{2(n-1)}{n-1}}
$$
\end{corollary}

\begin{proof}
$$
\begin{array}{rl}
\mu(d_T(i,j))_n & \displaystyle =\frac{\sum_{T\in \TT_n} d_T(i,j)}{|\TT_n|}
=\frac{S_n^{(1)}}{(2n-3)!!}=\frac{(2n-2)!!}{(2n-3)!!}\\[2ex] & \displaystyle =\frac{(2n-2)!!^2}{(2n-3)!!(2n-2)!!}=\frac{2^{2n-2}\cdot (n-1)!^2}{(2n-2)!}=\frac{2^{2(n-1)}}{\binom{2(n-1)}{n-1}}.
\end{array}
$$
\end{proof}

In the unrooted case, this mean value is proved in \cite{distributions} to be
${2^{2(n-2)}}/{\binom{2(n-2)}{n-2}}$.

\section{Mean path-difference distance}
The \emph{path-difference distance} between a pair of phylogenetic trees $T,T'\in \TT_n$ is
$$
\delta(T,T')=\sqrt{\sum_{1\leq i<j\leq n} (d_T(i,j)-d_{T'}(i,j))^2}.
$$

\begin{lemma}
The mean value of $\delta(T,T')^2$, with $T,T\in \TT_n$, under the uniform distribution on $\TT_n$, is
$$
\mu(\delta^2)_n=2\binom{n}{2}\left(4(n-1)+2-\frac{2^{2(n-1)}}{\binom{2(n-1)}{n-1}}-\left(\frac{2^{2(n-1)}}{\binom{2(n-1)}{n-1}}\right)^2\right).
$$
\end{lemma}

\begin{proof}
By definition
$$
\begin{array}{rl}
\mu(\delta^2)_n &  =\displaystyle \frac{\sum_{T,T'\in \TT_N}\sum_{1\leq i<j\leq n} (d_T(i,j)-d_{T'}(i,j))^2}{|\TT_N|^2}\\
&  =\displaystyle\frac{1}{|\TT_N|^2}\Big(\sum_{1\leq i<j\leq n}\sum_{T,T'\in \TT_N} (d_T(i,j)^2+d_{T'}(i,j)^2-2d_T(i,j)d_{T'}(i,j))\Big)
\\
&  =\displaystyle\frac{1}{|\TT_N|^2}\sum_{1\leq i<j\leq n}\Big(2|\TT_n|\sum_{T\in \TT_N} d_T(i,j)^2 -2\Big(\sum_{T\in \TT_N} d_T(i,j)\Big)^2\Big)
\end{array}
$$
and then, using Lemma \ref{lem:eq1},
$$
\begin{array}{rl}
\mu(\delta^2)_n&  =\displaystyle2\binom{n}{2}\left(\frac{\sum_{T\in \TT_n} d_T(1,2)^2}{|\TT_n|}-
\Big(\frac{\sum_{T\in \TT_n} d_T(1,2)}{|\TT_n|}\Big)^2\right)\\[1ex]
&  =\displaystyle 2\binom{n}{2} \left(\frac{S_n^{(2)}}{(2n-3)!!}-\Big(\frac{S_n^{(1)}}{(2n-3)!!}\Big)^2\right)
\end{array}
$$

If we replace $S_n^{(1)}$ and $S_n^{(2)}$ by their values given in Propositions \ref{prop:muD1} and \ref{prop:muD2}, we obtain
$$
\begin{array}{rl}
\mu(\delta^2)_n& \displaystyle =2\binom{n}{2}\left(\frac{2(2n-1)!!-(2n-2)!!}{(2n-3)!!}-\Big(\frac{(2n-2)!!}{(2n-3)!!}\Big)^2\right)\\
& \displaystyle 2\binom{n}{2}\left(4n-2-\frac{(2n-2)!!}{(2n-3)!!}-\Big(\frac{(2n-2)!!}{(2n-3)!!}\Big)^2\right)
\end{array}
$$
Applying finally (see Corollary \ref{cor:meand})
$$
\frac{(2n-2)!!}{(2n-3)!!}=\frac{2^{2(n-1)}}{\binom{2(n-1)}{n-1}},
$$
we obtain the expressions in the statement.
\end{proof}

The value of $\mu(\delta^2)_n$ obtained by Steel and Penny in the unrooted case 
was
$$
\mu(\delta^2)_n=2\binom{n}{2}\left(4(n-2)+2-\frac{2^{2(n-2)}}{\binom{2(n-2)}{n-2}}-\left(\frac{2^{2(n-2)}}{\binom{2(n-2)}{n-2}}\right)^2\right).
$$
Using  Stirling approximation, both mean values are equivalent to 
$$
2\binom{n}{2}((4-\pi)n-\sqrt{\pi n}).
$$

\section*{Acknowledgment}

The research reported in this paper has been partially supported by
the Spanish government and the UE FEDER program
project MTM2006-07773 COMGRIO. We thank G. Valiente for several comments on this work.

%
%\bibliographystyle{elsart-num}
%\bibliography{transdist}

\end{document}